\documentclass[aps,floats,superscriptaddress,showpacs]{revtex4}
\usepackage{amsmath}

\usepackage{epsfig}
\usepackage{graphicx}
\usepackage{dcolumn}
\usepackage{bm}
\usepackage{latexsym}

\def\gapp{\lower.35em\hbox{$\stackrel{\textstyle>}{\sim}$}}
\def\lapp{\lower.35em\hbox{$\stackrel{\textstyle<}{\sim}$}}

\begin{document}
\bibliographystyle{apsrev}

\title{Effects of topological defects and local curvature on the electronic
properties of planar graphene}
\author{Alberto Cortijo}
\affiliation{Unidad Asociada CSIC-UC3M,
Instituto de Ciencia de Materiales de Madrid,\\
CSIC, Cantoblanco, E-28049 Madrid, Spain.}

\author{ Mar\'{\i}a A. H. Vozmediano}
\affiliation{Unidad Asociada CSIC-UC3M,
Universidad Carlos III de Madrid, E-28911
Legan\'es, Madrid, Spain. }

\date{\today}
\begin{abstract}
A formalism is proposed to study
the electronic and transport properties of  graphene
sheets with corrugations as the one recently synthesized. The formalism is based
on coupling the Dirac equation that models the low energy electronic
excitations of clean flat graphene samples to a curved space.
A cosmic string analogy allows to
treat an arbitrary number of topological defects located at arbitrary
positions on the graphene plane. The usual defects that will always be present
in any graphene sample as pentagon-heptagon pairs and Stone-Wales defects are
studied as an example. The local density of states around the defects acquires
characteristic modulations
that could be observed in scanning tunnel and
transmission electron microscopy.

\end{abstract}
%

\maketitle
\section{Introduction.}
The recent synthesis of single layers of graphite
and the experimental confirmation of the properties predicted by
continuous models based on the Dirac equation\cite{Netal05,Zetal05}
have renew the interest in this type of materials.
Under a theoretical point of view, graphene has received a lot of attention in the past
because it constitutes a beautiful and simple  model of
correlated electrons in two dimensions
with unexpected physical properties\cite{KEetal03}. A tight-binding
method applied to the honeycomb lattice allows to describe the low energy
electronic excitations of the system around the Fermi {\rm points}
by the massless Dirac equation in two
dimensions. The density of states turns out to be zero at the Fermi points
making useless most of the phenomenological expressions
for transport properties in Fermi liquids. Among the unexpected properties
are the anomalous behavior of the quasiparticles decaying
linearly with frequency\cite{GGV96},
and the so-called axial anomaly
\cite{S84,Hal88} that has acquired special relevance in relation with
the recently measured anomalous Hall effect in graphene\cite{KEK06,Netal05,Zetal05}.

Disorder plays a very important role in the electronic
properties of low dimensional materials.  In graphene and fullerenes
the effect is even more drastic due to the vanishing
of the density of states at the Fermi level.
It is also an essential ingredient to search for the elusive magnetic behavior
\cite{Eetal03}. The influence of disorder on the electronic
properties of graphene has been intensely studied
recently. The classical works of disordered  systems described by
two dimensional Dirac fermions\cite{Letal94,NTW94,NTW95,SGV05} has been supplemented with
an analysis of vacancies, edges and cracks\cite{PGC06}.

Substitution of an hexagon by other type of polygon in the lattice
without affecting the threefold coordination of the carbon atoms leads to the warping
of the graphene sheet and is responsible for the formation of fullerenes.
These defects can be seen as disclinations of
the lattice which acquires locally a finite curvature. The accumulations of various
defects may lead to closed shapes. Rings with $n<6$ sides give rise to positively
curved structures, the most popular being the $C_{60}$ molecule that has twelve pentagons.
Polygons with $n>6$ sides lead to negative curvature as occur at the joining
part of carbon nanotubes of different radius and in the Schwarzite\cite{Petal03},
a structure appearing in many forms of carbon nanofoam\cite{Retal04}.
This type of defects have been observed in experiments
with carbon nanoparticles\cite{Hetal04,Getal94,Ketal97} and other layered materials\cite{TTM01}.
Conical defects with an arbitrary opening angle
can be produced by accumulation of pentagons in the
cone tip and have been observed in \cite{JRDG03,Anetal01}.
Inclusion of an equal number of pentagons and heptagonal rings in a graphene
sheet would keep the flatness of the sheet at large scales and produce
a flat structure with curved portions that would be structurally stable
and have distinct electronic properties.
This lattice distortions give rise to long range modifications
in the electronic wave function. The change of the local
electronic structure induced by a disclination is then very different
from that produced by a vacancy or other impurities modelled by local potentials.

In this  work we propose a model to study
the electronic properties of a graphene sheet with an arbitrary
number of topological defects that produce locally positive or negative
curvature to the graphene sheet. We will first perform a complete
description of the effect of disclinations on the low energy excitations
of graphene, write down the most general model, and solve it to find
the corrections to the density of states induced by heptagon-pentagon pairs
and Stone-Wales defects.

Disclinations can be included in the continuous model   as
topological vortices coupled to electronic excitations. We will
show that certain types of disclinations  produce a
non-vanishing local density of states at Fermi level. In the average flat
sheet of slightly curved graphene, heptagon-pentagon pairs can be described
as bound in dislocations that change the electronic properties of the
system.
The electronic properties induced by single
defects depend crucially on the
nature of the substitutional polygon.
Topological defects that involve the exchange of the Fermi points
(substitution of an hexagon by an odd-membered ring)
can break the electron-hole symmetry of the system
and enhance the local density of states that remains zero at the Fermi level.
The situation is similar to the effects found in the study of vacancies
in the tight binding model
when next to nearest neighbors (t') are included\cite{PGC06}.
Defects involving even-membered rings induce a non zero density of states
at the Fermi level preserving the electron-hole symmetry. An arbitrary
number of heptagon-pentagon pairs produce characteristic patterns
in the local density of states that can be observed in
scanning tunnel (STM)\cite{Aetal01} and
electron transmission spectroscopy (ETS).
The results obtained can help to interpret
recent Electrostatic Force Microscopy (EFM) measurements
that indicate large potential differences
between micrometer large regions on the surface of highly oriented graphite\cite{GE06}.

The rest of the paper is organized as follows: in Sect. 2 we review
briefly the main features of the continuous  model of graphene
based on the Dirac equation. We make special emphasis on the internal symmetries
that will be affected by the inclusion of topological defects.
In Sect. 3 single disclinations are introduced
in the model by means of gauge fields as a warmup exercise and as a way
to show the limitations of the model.
Substitution of an hexagon by
an even-membered ring is shown to induce  a finite density of states at the Fermi
level. Section 4 contains the main results of the paper.  A formalism is presented that
permits to study an arbitrary number of defects located at given positions in the
graphene lattice. The model is based on the observation that
the effect of a cosmic string on the space-time is the same
as the one produced by a pentagon in the two-dimensional
graphene plane. We generalize the cosmic string formalism
to include the effects of defects with an "excess angle" such
as heptagons and propose a metric to describe an arbitrary
number of disclinations in the graphene plane. The
electronic properties of the model are obtained from
the Green´s function of the system in the given metric.
We then apply the method to study the type of defects that are
most probably present in graphene samples: heptagon-pentagon pairs
and Stone-Wales defects. The main results are shown in section 5. We
show the inhomogeneous structures produced
in the density of states by these defects and argue that  they can
be observed in STM experiments. The last section contains the conclusions
and open problems.
Appendices A and B contain the technical details of the calculations of
sections 3 and 4 respectively.

\section{Low energy description of graphene. }

The conduction band of graphene is well described by a tight
binding model which includes the $\pi$ orbitals
which are perpendicular to the plane at each
C atom\cite{W47,SW58}. This model describes a semimetal, with zero
density of states at the Fermi energy, and where the
Fermi surface is reduced to two inequivalent K-points
located at the corner of the hexagonal
Brillouin Zone.

The low-energy excitations with momenta
in the vicinity of any of the Fermi points $K_+$ and $K_-$
have a linear dispersion
and can be described by a continuous model which reduces to the
Dirac equation in two dimensions\cite{GGV92,GGV93,GGV94}.
In the absence of interactions or disorder mixing  the two
Fermi points, the electronic properties of the system are well described
by the effective low-energy Hamiltonian density:
\begin{align}
{\cal H}_{0i}= \hbar v_{\rm F} \int d^2 {\bf r} \bar{\Psi}_i({\bf r})
( i \sigma_x \partial_x + i \sigma_y \partial_y )
\Psi_i ({\bf r})\;,\label{hamil}
\end{align}
where $\sigma_{x,y}$ are the Pauli matrices,
$v_{\rm F} = (3 t a )/2 $, and $a=  1.4\AA$ is the
distance between nearest carbon atoms. The
components of the two-dimensional wavefunction:
\begin{equation}
\Psi_i( {\bf r} )= \left( \begin{array}{c}
\varphi_A ( {\bf r} ) \\  \varphi_B ( {\bf r }) \end{array} \right)
\label{2spinor}
\end{equation}
correspond to the amplitude of the wave function in each of the two
sublattices (A and B) which build up the honeycomb structure.
We will later show that
the parameter space where this spinorial degree of freedom acts is
the polar angle of the real space of the graphene plane.
The dispersion relation
$\epsilon({\bf k})=v_F\vert {\bf k}\vert$
gives rise to the density of states
$$\rho(\omega)=\frac{8}{v_F^2}\;\vert\omega\vert$$
which vanishes at the Fermi level $\omega =0$.
The electronic states attached to the two inequivalent Fermi points
will be independent in the absence of interactions that
mix the two points.

The type of defects that we will study affect the microscopic  description  of graphene
in all possible ways: induce local curvature to the sheet, can mix
the two triangular sublattices, and can exchange the two Fermi points.
It is then convenient to set a unified description
and combine the bispinor attached to each Fermi point
(what is called in semiconductor´s language the valley degeneracy)
into  a four component Dirac spinor. We will do that and then analyze the
behavior of these pseudospinors under rotations what will be
crucial in the study of the boundary conditions imposed by the
defects.

The four dimensional Hamiltonian is
\begin{eqnarray}
H_{D}=-iv_{F}\hbar(1 \otimes \sigma_{1} \partial_{x}+\tau^{3}
\otimes \sigma_{2} \partial_{y}) ,\label{hamil4}
\end{eqnarray}
where $\sigma$ and  $\tau$ matrices are Pauli matrices acting on the
sublattice and valley degree of freedom respectively.
The dispersion relation associated to (\ref{hamil4}) is
\begin{eqnarray}
E(\textbf{p})=\pm\hbar v_{F}|\textbf{p}|\equiv \pm\hbar v_{F}p.
\label{dispersion}
\end{eqnarray}
The solutions of the Dirac equation - with positive energy - are of the form
\begin{eqnarray}
\Psi_{E>0}=exp(i\textbf{pr})\left(\begin{array}{c}
              e^{-i\theta/2} \\
             e^{i\theta/2} \\
             e^{i\theta/2} \\
              e^{-i\theta/2} \\
           \end{array}\right),
\label{4spinor}
\end{eqnarray}
where $\theta$ is the polar angle of the vector ${\bf p}$ in real space. The
first (second) two components of (\ref{4spinor}) refer to the bispinor around $K_+$
( $K_-$).

The behavior of (\ref{4spinor})  under a real space rotation of angle $\alpha$ around the
$oz$ axis is

\begin{eqnarray}
\Psi'(\textbf{r}')=\Psi'(R^{-1}\textbf{r})\equiv
T_{R}\Psi(R^{-1}\textbf{r}). \label{espinor_rot}
\end{eqnarray}

\noindent
the transformation
$\textbf{p}'=\textbf{p}R=p(cos(\alpha+\theta),sin(\alpha+\theta))$,
determines the $T_R$ matrix  to be

\begin{eqnarray}
T_{R}=\left(%
\begin{array}{cc}
  exp(i\frac{\alpha}{2} \sigma_{3}) & 0 \\
  0 & exp(-i\frac{\alpha}{2} \sigma_{3}) \\
\end{array}%
\right), \label{T_rot}
\end{eqnarray}

\noindent
what shows that (\ref{4spinor}) transforms as a real spinor under spacial
rotations of the graphene plane. Each of the two-dimensional K-spinor
transform under the given rotation with the matrix $\pm\sigma_3/2$.
This opposite sign is often referred to as the K-spinors having
opposite chirality
or helicity.

\section{Effect of a single disclination}

\begin{figure}
  \begin{center}
    \epsfig{file=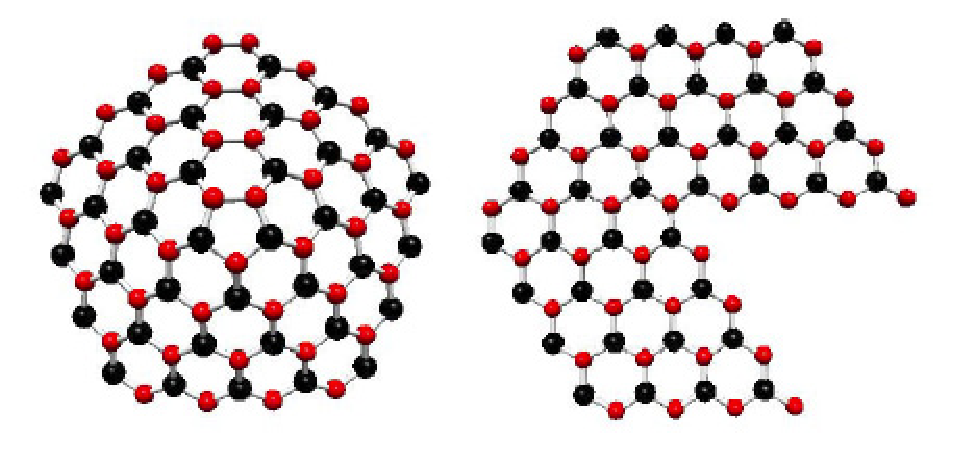,height=4cm}
    \caption{Left: Effect of a pentagonal defect in a graphene layer.
    Right: Cut-and-paste procedure to form the pentagonal defect. The  points at the
    edges are connected by a link what induces a frustration of the bipartite character of
    the lattice at the seam.
    }
    \label{fig1}
\end{center}
\end{figure}

Substitution of an hexagon by an n-sided
polygon in the graphene lattice can be described by a cut-and-paste procedure
as the one shown in fig. \ref{fig1}  for the particular case of a pentagon.
A $\pi/3$ sector of the lattice is removed and the edges are glued.
In this case the planar lattice acquires the form of a cone with the pentagon in
its apex.
Such a disclination has two distinct effects on the
graphene sheet. It induces locally positive (negative) curvature
for $n<6$ ($n>6$) and, in the paste procedure, it
can break the bipartite nature of the lattice if n is odd
while preserving the symmetry if n is even. This makes a difference
with the case of the formation of nanotubes where the bipartite
nature of the lattice always remains intact.
The presence of an odd-membered ring means that
the two fermion flavors defined in eq. (\ref{2spinor}) are also exchanged
when moving around such a defect\cite{GGV92}. The
scheme to incorporate this change in a continuous
description was discussed in refs. \cite{GGV93} and \cite{GGV01}.
The process can be described by means
of a non Abelian gauge field, which rotates the spinors in
flavor space. As will be shown the two cases have very different effect on the
density of states of the system.
Conical defects with an arbitrary opening angle
made by accumulation of pentagons at the tip
have been observed experimentally in\cite{JRDG03}.

We shall begin describing the effect on the density of states
produced by conical defects that
do not alter the bipartite character of the hexagonal lattice.

A spinor defined in a plane without topological defects
acquires a phase of $\pi$ (changes sing) when going around
a closed path\cite{Ryder}. In the two-dimensional
description this can be written as
$$\psi_0({\bf r},\varphi=2\pi)=e^{i\pi\sigma_3}\psi_0({\bf r},\varphi=0).
\label{S0}$$
When the spinor rotates around a defect with a deficit angle $b=2\pi b$ it
obeys the boundary condition
$$\psi({\bf r},\varphi=2\pi)=e^{i2\pi(1-b)
\frac{\sigma_3}{2}}\psi({\bf r},\varphi=0).
$$
We can  convert the phase $b$ in a continuous
variable and assume that a lattice distortion which rotates the lattice axis
can be parametrized by the angle of rotation, $\theta ({\bf \vec{r}} )$,
of the local axes with respect to a fixed reference frame. The spinor
can be written as
\begin{equation}
\psi({\bf r},\varphi)=e^{i(\int_x{\bf A}({\bf y}){\bf dy})\frac{\sigma_3}{2}}
\psi_0({\bf r},\varphi),
\label{spinor2}
\end{equation}
with
\begin{equation}
 {\bf A}( {\bf r} )  \sim\nabla \theta ( {\bf r} ).
\nonumber
\end{equation}
In the four-dimensional representation, applying the
Dirac operator $i\gamma . \nabla$ to eq. (\ref{spinor2})
we get the following hamiltonian:
\begin{eqnarray}
H=-i\hbar
v_{_{F}}{\vec \gamma}.{\vec\partial}+g\gamma^{q}
{\vec \gamma}.{\vec A}({\bf r}),
\label{hamgauge}
\end{eqnarray}
where $v_{_{F}}$ is the Fermi velocity,
$ \gamma^{i}$ are
$4\times4$ matrices constructed from the Pauli matrices,
$\gamma^q=\frac{\sigma_3}{2}\otimes I$,
the latin
indices run over the two spatial dimensions and g is a coupling
parameter.
 The external field $A_{i}({\vec{r}})$ takes the form of a
vortex
\begin{eqnarray}
A^{j}({\vec{r}})=
\frac{\Phi}{2\pi}\epsilon^{3ji}\frac{x_{i}}{r^{2}}\label{gaugefield}
\end{eqnarray}
The constant $\Phi$ is a parameter that represents the strength of
the vortex: $\Phi=\oint\vec{A}d\vec{r} $ and is related to the opening
angle of the defect. A geometric formulation
of the same problem has been given recently in \cite{FMC06}
in terms of holonomy.
In the case of an odd-membered ring, the two Fermi points
are also exchanged\cite{GGV93} what can be modelled by
using a non-abelian  gauge potential that rotates the spinors
in the $SU(2)$ space of the Fermi points. In this case an extra matrix
appears in the coupling of the gauge field in (\ref{hamgauge}).
In what follows we will consider
the static vortex as an external gauge potential. This approximation
is justified by the fact that
the dynamics of the defects in the
lattice is related to the $\sigma$ bonds
with energy of about 4 eV whereas the
electronic excitations described by the spinors
involve energies of the order of 20 meV.
We will then use time-independent
perturbation theory to calculate corrections to the self-energy
$\Sigma(\textbf{k},\omega)$
in the weak coupling regime of the parameter
$$\widehat{g}\equiv \frac{\Phi}{2\pi L^2},\label{singlecoupling}$$
where L is the dimension of the sample.
The correction to the density of states
induced by the defect is obtained from the self-energy by
\begin{eqnarray}
\rho(\omega)=\frac{1}{\pi}Im\int Tr G(\omega, {\bf k}) \nonumber
\end{eqnarray}
The presence of the defect breaks the translational invariance and
the computation of the density of states deviates slightly from
the standard path. The details of the calculation are given in appendix A.
\begin{figure}
  \begin{center}
    \epsfig{file=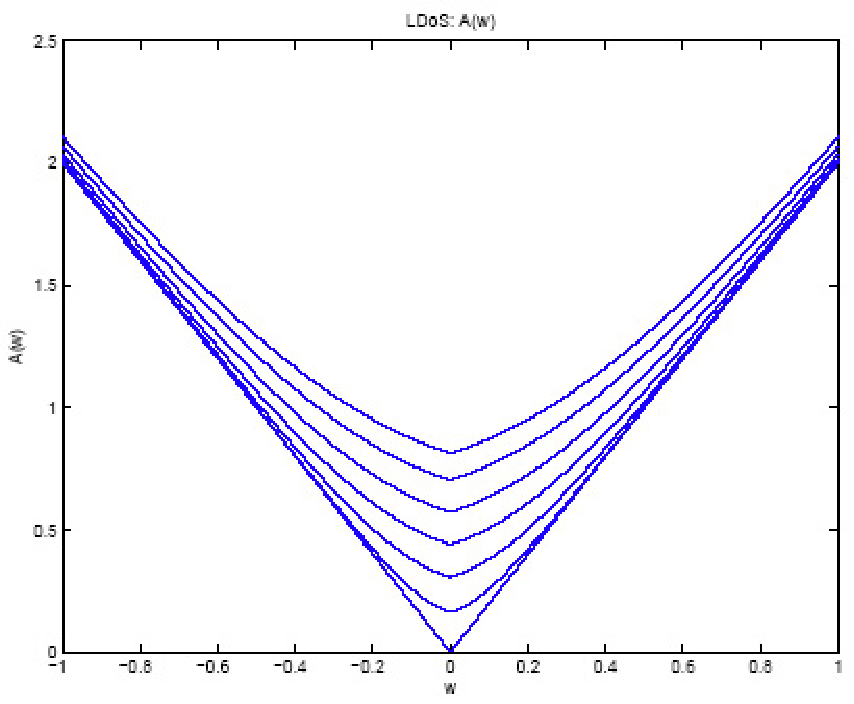,height=4cm}
    \caption{Total density of states  for an even-membered ring
    for several increasing values of
    $\widehat{g}$ (see text) in arbitrary units starting with $\widehat{g}$=0.
    }
    \label{doseven}
\end{center}
\end{figure}
\begin{figure}
  \begin{center}
    \epsfig{file=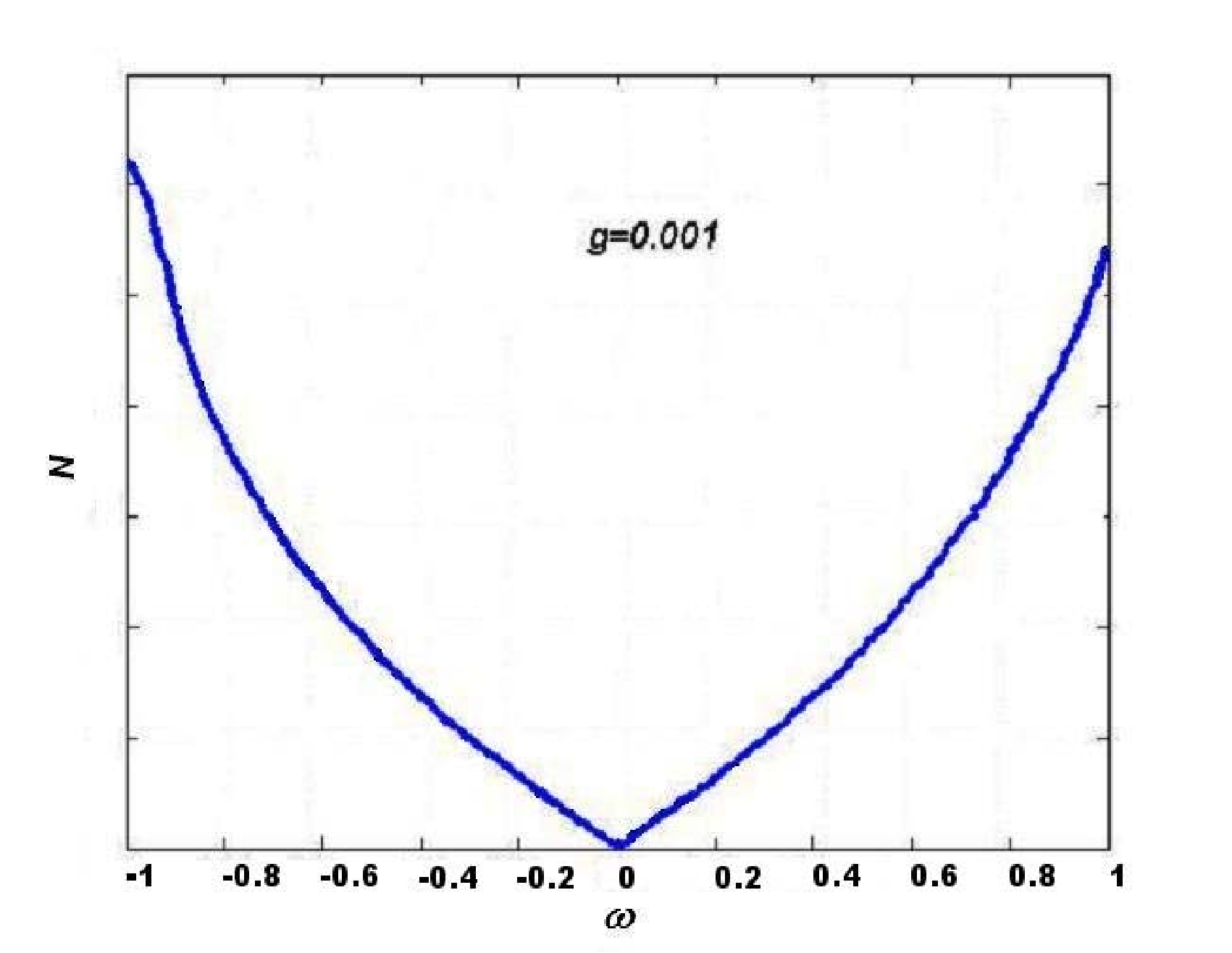,height=4cm}
    \caption{Total density of states  for an odd-membered ring.
    }
    \label{dosodd}
\end{center}
\end{figure}

The correction to the density of states given in (\ref{DOS}) is shown
in Fig. \ref{doseven} for several values of the
coupling parameter $\widehat{g}$. We see that the defect
induces a non-zero density of states at the Fermi
energy  given by
\begin{equation}
\rho(\omega=\varepsilon_F,\mid\widehat{g}\mid)=
\frac{2\sqrt{\mid\widehat{g}\mid}}{v_F}.
\label{dosF}
\end{equation}
The value of the DOS increases with the curvature (encoded
in the parameter $\widehat{g}$). It depends on the size of the
sample and will go to zero as (1/L) in the thermodynamic limit.
 Similar results were obtained
in \cite{LC00} and \cite{OK01} where the same problem is
addressed but there they compute the local DOS around a
defect truncating the singularity at the apex of the cone.
Numerical ab initio calculations show sharp resonant peaks  in the LDOS at
the tip apex of nanocones \cite{TT94,CR01}
that have been proposed for
electronic applications in field emission devices. In both cases
the computations refer to the local density of states.

The case of an odd membered ring is technically more involved
in our formalism and we have not obtained an analytical result.
The complication is produced by the extra matrix
needed to exchange the two fermion flavors. A numerical integration
of the DOS for this case is shown in
fig. \ref{dosodd} for the case of a single pentagonal defect.
We can see that the DOS at the Fermi level is zero in this
case also in agreement with \cite{LC00} and \cite{OK01}.  We
can also appreciate a
small deviation from the perfect electron-hole symmetry of $\rho(\omega)$.
The slope of the curve at the origin suggests an enhancement of the DOS
around the zero energy what would agree with the STM observations
described in \cite{Anetal01}.
We will come back to the single defect in the next section.

\section{An arbitrary number of defects at given positions in the lattice.}
An alternative approach to the gauge theory of defects discussed
in the previous section is to include the local curvature induced by an n-membered ring
by coupling the Dirac equation to a curved space.
In this context  one can see that the substitution of
an hexagon by a polygon of $n<6$ sides
gives rise to a conical singularity with deficit angle
$(2\pi/6)(6-n)$. This kind of singularities have been studied
in cosmology as they are produced by  cosmic strings, a type of
topological defect that arises when a U(1) gauge symmetry is
spontaneously broken\cite{vilenkin}.
We can obtain the correction to the density of states
induced by a set of defects with arbitrary opening angle
by coupling the Dirac equation to a curved space with an appropriate
metric as described in ref. \cite{AHO97}.
In this section we will closely
follow the formalism set in these reference.

The metric of a two dimensional space in presence of a single cosmic
string in polar coordinates is:
\begin{equation}
ds^{2}=-dt^{2}+dr^{2}+c^{2}r^{2}d\theta^{2},\label{metric}
\end{equation}
where the parameter $c$ is a constant related to the deficit angle
by $c=1-b$.

The dynamics of a massless Dirac spinor in a curved spacetime is
governed by the Dirac equation:
\begin{equation}
i\gamma^{\mu}\nabla_{\mu}\psi=0 \label{dircurv}
\end{equation}
The difference with the flat space  lies in the definition
of the $\gamma$ matrices that satisfy generalized anticommutation
relations
\begin{equation}
\{\gamma^{\mu},\gamma^{\nu}\}=2g^{\mu\nu},\nonumber
\end{equation}
and in the covariant derivative operator, defined as
$$
\nabla_{\mu}=\partial_{\mu}-\Gamma_{\mu}
$$
where $\Gamma_{\mu}$ is the spin connection of the
spinor field that can be
calculated using the tetrad formalism\cite{birrell}.
\begin{figure}
  \begin{center}
    \epsfig{file=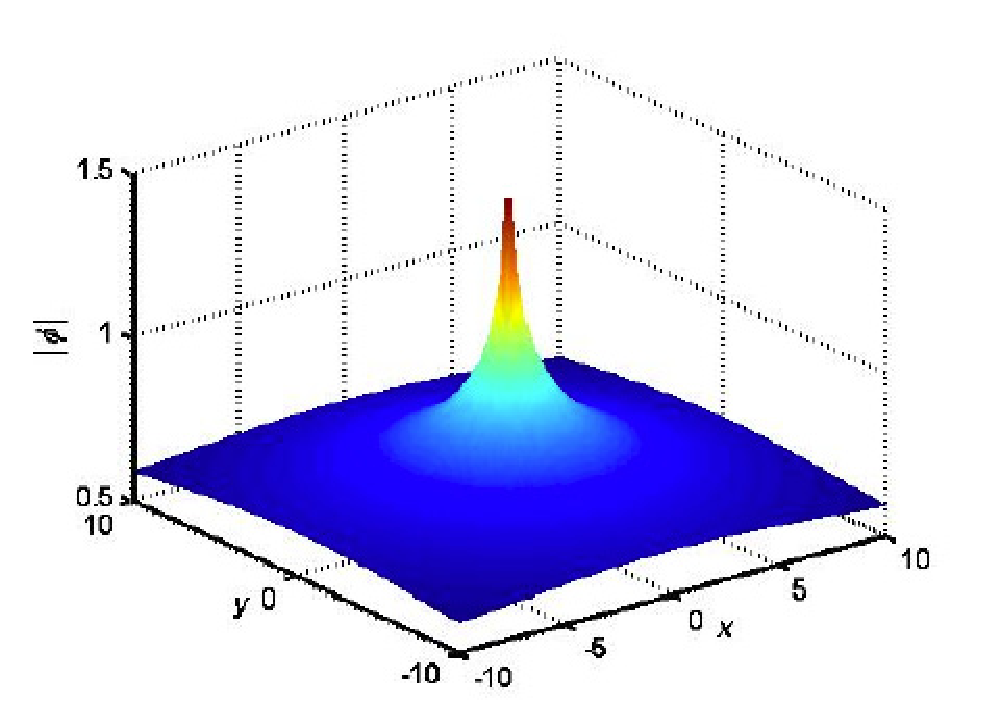,height=6cm,width=6cm}
    \caption{Electronic density  around a conical defect.
    }
    \label{1string}
\end{center}
\end{figure}
The present formalism can help to clarify the nature of the
states appearing at the Fermi level in fig. \ref{doseven}.
Fig. \ref{1string} shows the solution of the Dirac
equation (\ref{dircurv}) in the presence
of a single defect with a positive deficit angle
(positive curvature). The electronic density is strongly peaked at the
position of the defect suggesting a bound state but
the behavior at large distances is a power law with an angle-dependent
exponent less than two
which corresponds to a non normalizable wave function. This  behavior  is similar to the
one found in the case of a single vacancy\cite{PGLPC06}
and suggests that a
system with a number of this defects with overlapping
wave functions will be metallic.

The case of   a single cosmic string which
represents a deficit angle in the space can be generalized
to describe seven membered rings representing an angle
surplus by considering a value for c
larger than $1$.  This situation is non-physical from a general
relativity viewpoint as it would correspond to a string with negative mass
density but it makes perfect sense in our case. The scenario can also
be generalized to describe an arbitrary number of pentagons and
heptagons by using  the following metric:
\begin{equation}
ds^{2}=-dt^{2}+e^{-2\Lambda(x,y)}(dx^{2}+dy^{2}),\label{genmetric}
\end{equation}
where $$\Lambda(\textbf{r})=\sum^{N}_{i=1}4\mu_{i}\log(r_{i})$$ and
$$r_{i}=[(x-a_{i})^{2}+(y-b_{i})^{2}]^{1/2}.$$ This metric describes
the space-time around N parallel cosmic strings, located at the
points $(a_{i},b_{i})$. The parameters $\mu_{i}$ are related to the
angle defect or surplus by the relationship $c_{i}=1-4\mu_{i}$ in
such manner that if $c_{i}<1 (>1)$ then $\mu_{i}>0 (<0)$.

From equation (\ref{dircurv}) we can
write down
the Dirac equation for the electron propagator, $S_{F}(x,x')$:
\begin{equation}
i\gamma^{\mu}({\bf r})(\partial_{\mu}-\Gamma_{\mu})S_{F}(x,x')=
\frac{1}{\sqrt{-g}}\;\delta^{3}(x-x'),
\label{propcurv}
\end{equation}
where $x=(t, {\bf r})$.
The local density of states $N(\omega,\textbf{r})$ is obtained from (\ref{propcurv})
by Fourier transforming the time component and taking the limit
${\bf r}\to {\bf r'}$:
\begin{equation}
N(\omega,\textbf{r})=Im
Tr S_{F}(\omega,\textbf{r},\textbf{r}).\label{LDOS}
\end{equation}
We solve eq. (\ref{LDOS}) considering the  curvature induced
by the defects as a perturbation of the flat graphene layer.
The details of the calculation are given in  Appendix B.
Here we will show the results obtained.

We must notice that the present formalism takes into account
the effects produced by the local curvature
of the lattice but does not
include yet the effect
of identifying points of different sublattices. It
is then specially suitable  to describe pentagon-heptagon
pairs or Stone-Wales defects where the effect of
the line of dislocation is minimized.

\section{The local density of states.}
\begin{figure}
  \begin{center}
    \epsfig{file=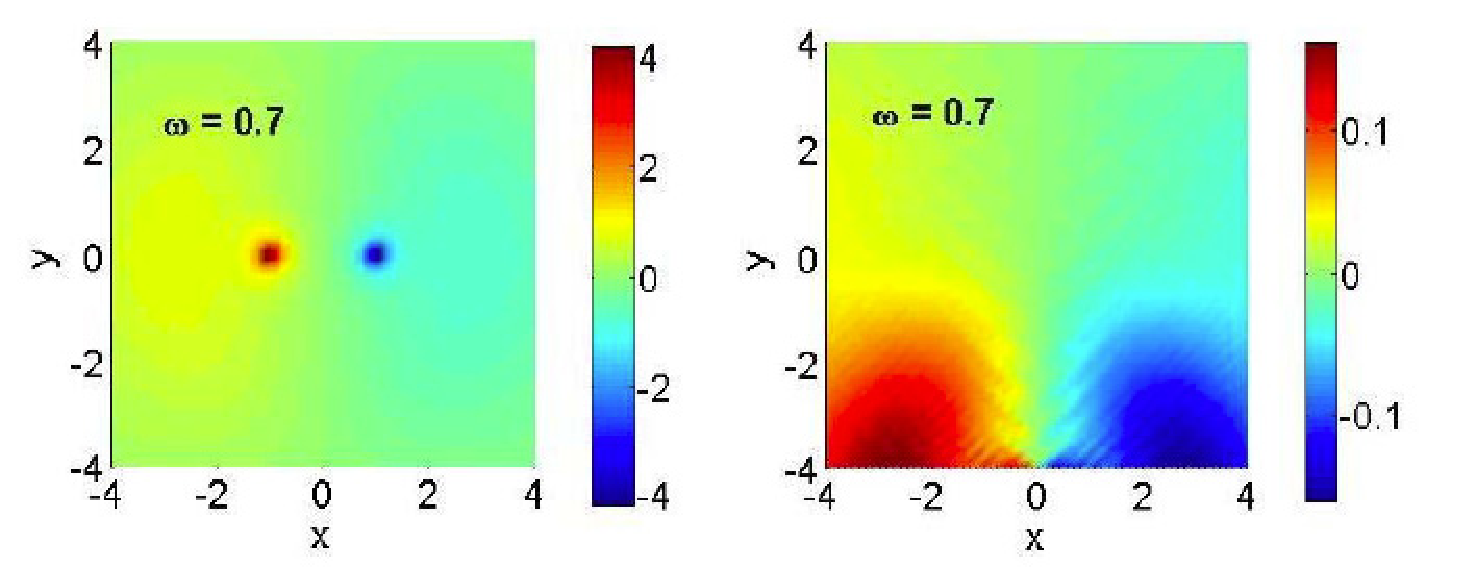,height=5cm}
    \caption{Left: Image of the local density of states in a large portion
    of the plane with a heptagon-pentagon  pair located at the center.
    Green color represents the DOS of the flat graphene sheet.
    Right: Same with the defect located out of plane.}
    \label{pair}
\end{center}
\end{figure}

In this section we will show the results obtained by applying the
cosmic string formalism to various cases of physical interest.
The left side of Fig. \ref{pair} shows the correction to the local density of states
at fixed energy and for a large region of the graphene plane with a
pentagon-heptagon pair in the middle. The color code is indicated in
the figure: green stands for the DOS of perfect graphene at the given energy
and red (blue) indicates an accumulation (depletion) of the density in the
area. We can see that pentagonal (heptagonal) rings  enhance
(deppress) the electron density. A similar result has been obtained in
\cite{TT94} with numerical simulations. It is to note that a somehow
contradictory result was obtained in \cite{AFM98} where they studied
the electrostatics of a graphene plane with defects. They found that
disclinations corresponding to rings with more (less) than six carbon atoms function
as attractors (repellent) to point charges. In the latter approach
they were concerned exclusively with the curvature
effect not taking into account the connectivity of the lattice. It is obvious that this
issue needs further investigation.
The right han side of Fig. \ref{pair} represents the structure of the density
of states produced by the same defect
located out of the plane
in the lower part. Notice the different intensities in the two graphics.
The dipolar character of the defect is clear.
%
\begin{figure}
  \begin{center}
    \epsfig{file=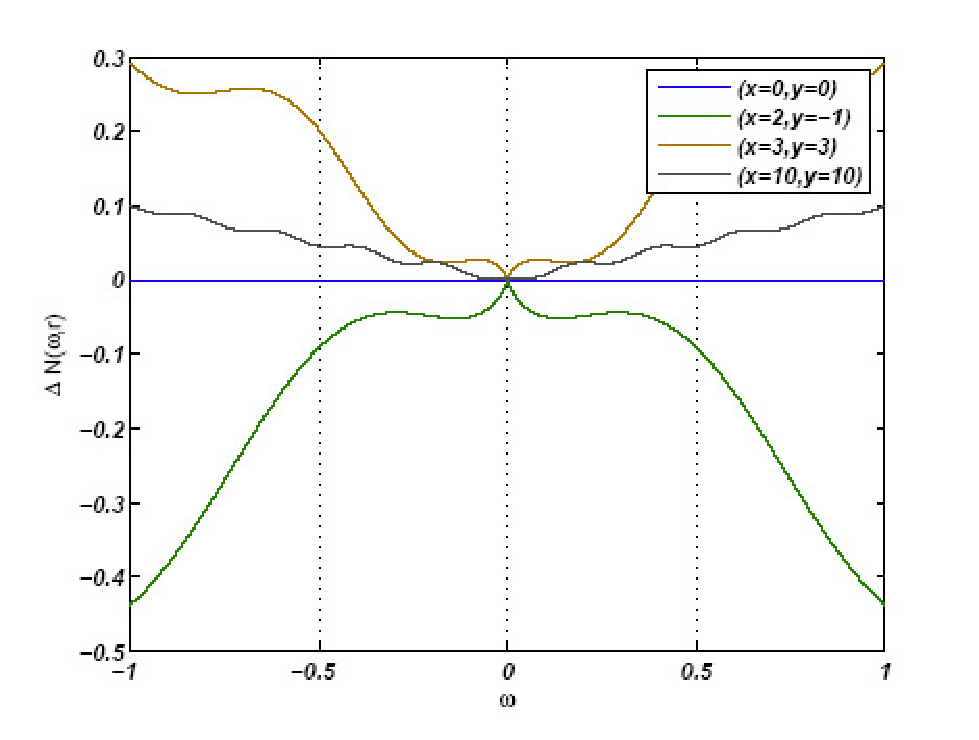,height=5cm}
    \caption{Correction  to the local density of states for
    different points of the lattice from a  pentagon-heptagon pair
    located at $(x=0,y=\pm 1)$.
    }
    \label{corrpair}
\end{center}
\end{figure}
Fig. \ref{corrpair} illustrates the same phenomena in different
coordinates. We show the  contribution to the local density of states
(LDOS) as a function of the energy
coming from a pentagon-heptagon pair located at $(x=0,y=\pm 1)$ computed at
different points of the plane. We can see that
the LDOS shows oscillations depending on the position of the
point relative to the position of the defect. The correction is zero in the
line perpendicular  to the segment joining the two defects as can
be seen in fig. \ref{pair2}.

\begin{figure}
  \begin{center}
    \epsfig{file=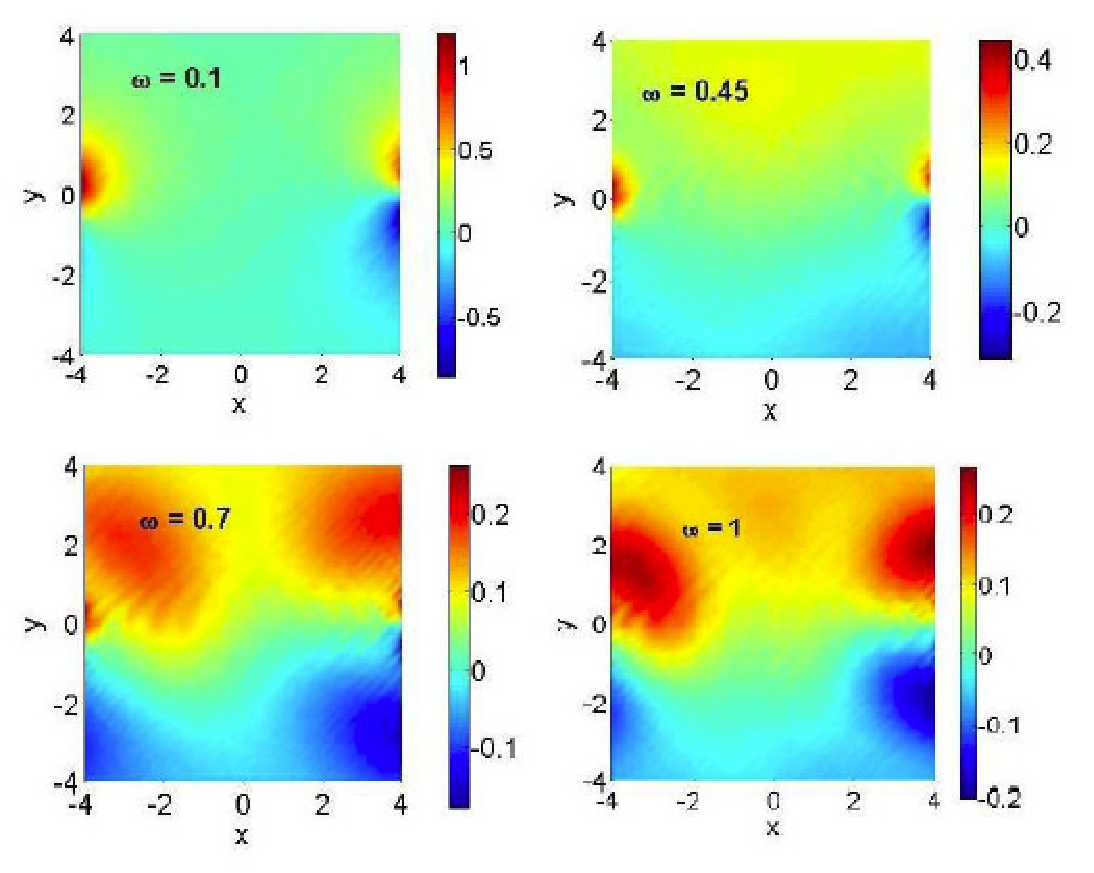,height=10cm}
    \caption{ Correction to the    local density of states in a wide region around two pairs
    of heptagon-pentagon defects located out of the region (see text)
    for increasing values of the energy.
    }
    \label{sequence}
\end{center}
\end{figure}

The intensity of the oscillations grows with the energy.
Fig. \ref{sequence} shows the correction to the local density of states in a
extended region of the lattice
induced by two pairs of heptagon-pentagon defects located out of the region
for increasing values of the energy. The first pair is located in the
down-left diagonal direction approximately at coordinates
(-4.5, -4.5) of figure and the second pair is located approximately
at (4.5,0).
In fig. \ref{sequence} a) in the space between  the defects the LDOS
is almost zero except at local zones where
the correction is small (in the region in the left down side
the density is enhanced by the proximity of the defects). As
the frequency increases up to $\omega=1$ near to the energy cutoff
(fig. \ref{sequence}c)
the LDOS is enhanced  and inhomogeneous
oscillations can be observed in the region between the defects.
The patterns depend also on the relative orientation of the dipoles.
These type of structures can be observed in experiments of scanning
tunnel spectroscopy.
In general the intensity of the oscillations grows with the energy.
Fig. \ref{sequence} shows the relative correction (normalized to the
free density of states) to the local density of states in a
extended region of the lattice
induced by two pairs of heptagon-pentagon defects located out of the region
for increasing values of the energy. The first pair is located in the
down-left diagonal direction approximately at coordinates
(-4.5, -4.5) of figure and the second pair is located approximately
at (4.5,0).
In fig. \ref{sequence} a) in the space between  the defects the LDOS
is almost zero except at local zones where
the correction is small (in the region in the left down side
the density is enhanced by the proximity of the defects). As
the frequency increases up to $\omega=1$ near to the energy cutoff
(fig. \ref{sequence}c)
the LDOS is enhanced  and inhomogeneous
oscillations can be observed in a wide area around the defects.
The patterns depend also on the relative orientation of the dipoles.
The spacial extent of the correction is such that the intensity
decays to ten percent in approximately 20 unit cells so they
can be observed in  scanning tunnel spectroscopy as
inhomogeneous regions of a few nanometers.

\begin{figure}
  \begin{center}
   \epsfig{file=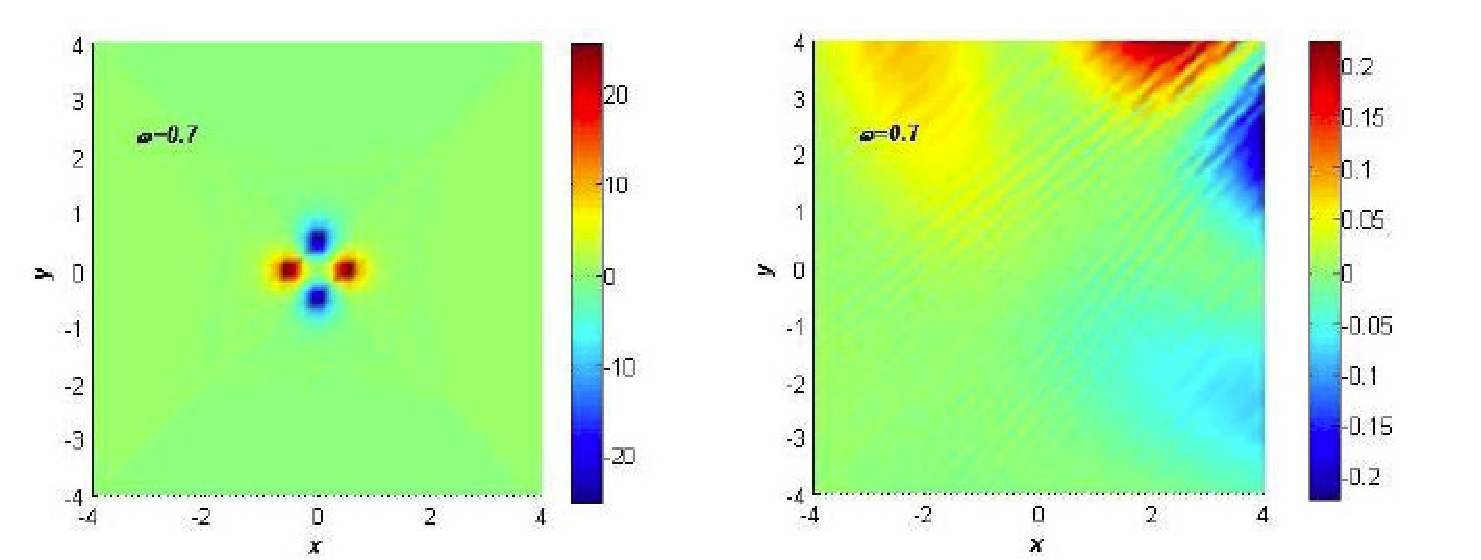,height=5cm}
   \vspace{3mm}
    \caption{Left: Local density of states around a Stone-Wales defect
    located at the center for a fixed value of the frequency. Right: same for a defect
    located out of plane (see text).}
    \label{stone-wales}
    \end{center}
\end{figure}

Stone-Wales defects made of two adjacent heptagons and two pentagons
are known to be energetically quasi-stable\cite{SW86} and
form naturally in experiments of ion bombarded nanotubes as a mechanism
to reduce the dangling bonds in large vacancies\cite{Aetal98}.
They have received a lot of attention
in carbon nanotubes where they strongly
affect the electronic properties\cite{CBLC96}. In graphene
they have been shown to change the band structure
causing a nondispersive spin-polarized band to form
and a peak in the density of states close to the Fermi level\cite{INFK96}.
They have also been shown recently to change substantially
the electronic structure of hydrogen
adsorbed in graphene\cite{DML04}.
In the left side of fig. \ref{stone-wales} we show an image of an extended region
of the graphene plane with a Stone-Wales defect located in the middle
at a fixed intermediate frequency. The modulation of the local
density of states around the defects is hardly noticeable due to the strong
intensity localized at the defects. The same image is shown in the right hand side of
fig. \ref{stone-wales} with the Stone-Wales defect located out of
plane in the upper right corner. The modulation of the LDOS is clearly
visible.

\section{Conclusions and open problems.}
We have presented a formalism based on cosmic strings to
study the electronic structure of a
slightly warped graphene sheet where the local curvature is produced by
substitution of an hexagon by an n-sided polygon with $n\neq 6$. This type
of disorder induces long range forces in the
graphene system and is very different from
the most studied cases of vacancies, cracks, edges, etc. We
have shown that
the local density of states is enhanced around defects with $n<6$ which induce
positive curvature in the lattice while the charge is
"repelled" from regions with negative curvature ($n>6$).
A single conical defect with an even number of
hexagonal sides excised produces a finite DOS
at the Fermi energy of a value that
increases with the curvature of the defect in agreement with STM
observations\cite{Anetal01}. The zero energy electronic states
are peaked at the defect but the wave function remains finite
at large distances so they  are extended states.
Heptagon-pentagon
pairs that keep the graphene sheet flat
in the long range behave as dipoles
and give rise to characteristic modulations of the DOS that can be observed
by STM. The same behavior is induced by Stone-Wales defects which are
expected to be produced by annealing of irradiated samples
as observed in carbon nanotubes\cite{Aetal98}. The magnitude
of the oscillations increases with the frequency and the characteristic patterns
could be used to characterize the graphene samples.
The features predicted in this work should also be observable in
other layered materials with similar structure as boron nitride\cite{TTM01}.
The present analysis can help to clarify the issue of the analysis
and interpretation of STM images.
\section{Acknowledgments.}
We thank  B. Valenzuela, P. Guinea and P. Esquinazi for many illuminating discussions.
A. C. thanks N. Ozdemir for kindly  providing details of the calculations
in ref. \cite{AHO97}.
Funding from MCyT (Spain) through grant MAT2002-0495-C02-01
and from the European Union Contract No. 12881 (NEST) is
acknowledged.

\appendix
\section{A single defect}
In a disordered system, in general, because of the presence of an
external potential or impurities, the space is inhomogeneous.
The Green's function doesn't depend on the difference ($\bf{r}-r'$)
and $\bf{k}$ is no longer a good quantum number.
If the external potential or the effect of the impurities
are time-independent we have elastic scattering and the states
\textbf{k} and \textbf{k}' have the same energy.

We want to calculate the total density of states
$\rho(\omega)$ of the system perturbed by the defect via the vector
potential given in eq. (\ref{gaugefield}). This density is the imaginary part of
the Green's function integrated over all positions, in the limit
$\textbf{r}'\rightarrow \textbf{r}$:
\begin{equation}
\rho(\omega)=\int Im
G(\omega,\textbf{r},\textbf{r})d\textbf{r}.\nonumber
\end{equation}
In terms of the Green's function in momentum representation,
$\rho(\omega)$ can be written as:
\begin{equation}
\rho(\omega)=Im\int\int \frac{d\textbf{k}}{(2\pi)^{2}} \int
\frac{d\textbf{k}'}{(2\pi)^{2}}
e^{i\textbf{kr}}e^{i\textbf{k}'\textbf{r}}G(\omega,\textbf{k},\textbf{k}')d\textbf{r}.
\nonumber
\end{equation}
The integration over $\textbf{r}$  gives delta function
$4\pi^{2}\delta(\textbf{k}+\textbf{k}')$, and $\rho(\omega)$ then
reads:
\begin{equation}
\rho(\omega)=\int \frac{d\textbf{k}}{(2\pi)^{2}}Im
G(\omega,\textbf{k},\textbf{-k}).\nonumber
\end{equation}
With this expression  we can compute ordinary Feynmann
diagrams contributing to the self-energy
$\Sigma(\omega,\textbf{k},\textbf{k}')$, calculate the Green's
function $G(\omega,\textbf{k},\textbf{k}')$ and make the
substitution $\textbf{k}'=-\textbf{k}$.
The first order correction to the electron self-energy is
\begin{eqnarray}
\Sigma(\textbf{k},\omega)=\Sigma_{0}(\textbf{k})+\widehat{g}
\gamma^{q}\gamma^{i}\langle
\textbf{k}|A_{i}(\vec{r})|\textbf{k}\rangle+O(g^{2})\label{selfen}
\end{eqnarray}
At this order the first contribution to the self-energy is simply
the Fourier transform of the gauge potential eq. (\ref{gaugefield})
and does not depend on
$\omega$. It turns out to be:
\begin{eqnarray}
\Sigma (\textbf{k},\omega)\simeq v_{_{F}}\gamma
^{i}k_{i}+\widehat{g} \gamma ^{i}\frac{k_{i}}{k^{2}}\equiv
v_{_{F}}f(k^{2})\gamma^{i}k_{i}, \label{selfen2}
\end{eqnarray}
where $f(k^{2})$ is
$f(k^{2})=1+\frac{\widehat{g}}{k^{2}}$. The two-point Green's function
reads, at this level:
\begin{eqnarray}
G(\omega, {\bf k})=\frac{w+f\gamma^{i}k_{i}}{\omega^{2}-f^{2}k^{2}+i\delta}\label{greens}
\end{eqnarray}
We compute the density of states of the system
\begin{eqnarray}
\rho(\omega)=Im\int Tr G(\omega, {\bf k})\nonumber
\end{eqnarray}
with the Green's function defined in eq. (\ref{greens}). The result is
\begin{eqnarray}
\rho(\omega)=\frac{4\omega
v^{4}_{_{F}}(K^{+}(\omega))^{4}}{|v^{4}_{_{F}}(K^{+}(\omega))^{4}
-\widehat{g}^{2}|},\label{DOS}
\end{eqnarray}
where
$K^{+}(\omega)=\sqrt{\frac{1}{2}(2\widehat{g}+v_{_{F}}
\omega^{2}+\sqrt{\omega^{4}+2\omega^{2}\widehat{g}
v_{_{F}}})}$.
\section{Multiple defects}
In the metric defined by (\ref{genmetric}):
\begin{equation}
g_{\mu\nu}=
\left(%
\begin{array}{ccc}
  -1 & 0 & 0 \\
  0 & e^{-2\Lambda} & 0 \\
  0 & 0 & e^{-2\Lambda} \\
\end{array}%
\right),
\end{equation}
the gamma matrices and the spinor
connection in the curved background   are found to be
$$
\gamma^{0}(\textbf{r})=\gamma^{0}\;,\;
\gamma^{i}(\textbf{r})=e^{\Lambda({\bf r})}\gamma^{i}\;\;
(i=1,2)
$$
$$
\Gamma_{1}(\textbf{r})=-\frac{1}{2}\gamma^{1}\gamma^{2}\partial_{y}\Lambda\;\;,\;\;
\Gamma_{2}(\textbf{r})=-\frac{1}{2}\gamma^{2}\gamma^{1}\partial_{x}\Lambda\;,
$$
and the determinant of the metric tensor is
$$
\sqrt{-g}=e^{-2\Lambda}.
$$
We solve eq. (\ref{propcurv}) by considering $\mu$ as a
small parameter to perturb around the flat space. In the case
that the defects are made  of pentagon
heptagon pairs the value of $\mu$ is
$\mu_{i}\equiv\mu=1/24$ $(b=1/6)$.

We expand  the function
$\Lambda(\textbf{r})$ to first order in $\mu$ and
get  for the electron propagator the equation
\begin{equation}
i\gamma^{0}\partial_{0}S_{F}-i\gamma^{j}\partial_{j}S_{F}-VS_{F}=\delta^{3}(x-x').
\label{propV}
\end{equation}
As before latin indices run over spatial dimensions.
We can see that eq. (\ref{propV}) is the equation for the Green´s
function of a spinor field in flat space but in an
an external potential $V$ given by:
\begin{equation}
V(\omega,\textbf{r})=2i\Lambda\gamma^{0}\partial_{0}+i\Lambda\gamma^{j}
\partial_{j}+\frac{i}{2}\gamma^{j}(\partial_{j}\Lambda).\label{effV}
\end{equation}
As in the case of a single defect, the effect of the curvature has been
traded to an external potential.
In order to get the corrections to the density of
states as a function of the frequency and position we
perform a time fourier transform in (\ref{propV}) and get an equation
for the quantity $\hat{S}_{F}(\omega,\textbf{r},\textbf{r}')$:
\begin{equation}
\gamma^{0}\omega\hat{S}_{F}-i\gamma^{j}\partial_{j}\hat{S}_{F}-
\hat{V}\hat{S}_{F}=\delta^{2}(\textbf{r}-\textbf{r}')\;,\label{prophat}
\end{equation}
where $\hat{V}$ is the time fourier transform of the potential
(\ref{effV}).
The first order correction to the
propagator in real space is
\begin{equation}
\hat{S}^{1}_{F}(\omega,\textbf{r},\textbf{r}')=\int d^{2}r''
\hat{S}^{0}_{F}(\omega,\textbf{r},\textbf{r}'')\hat{V}(\omega,\textbf{r}'')
\hat{S}^{0}_{F}(\omega,\textbf{r}'',\textbf{r}')\; \label{propw}
\end{equation}
where
\begin{equation}
\hat{S}^{0}_{F}(\omega,\textbf{r},\textbf{r}')=\int
\frac{d^2k}{(2\pi)^{2}}
\left(\frac{\gamma^{0}\omega-\gamma^{j}k_{j}}{\omega^{2}-k^{2}+i\delta}
\right)e^{-i\textbf{k}(\textbf{r}-\textbf{r}')} \label{free}
\end{equation}
is the free fermion propagator.
Writing  the function $\Lambda$ as
\begin{equation}
\Lambda(\textbf{r}'')=\int\frac{d^{2}p}{(2\pi)^{2}}e^{i\textbf{p}\textbf{r}''}
\Lambda(\textbf{p}),\nonumber
\end{equation}
we can use (\ref{free}) (\ref{propw}) and integrate out the
$\textbf{r}''$ variable. We then take the
limit $\textbf{r}'\rightarrow\textbf{r}$ and  the trace of
(\ref{propw}) and get
\begin{eqnarray}
Tr\hat{S}^{1}_{F}(\omega,\textbf{r},\textbf{r})=\frac{\mu}{2\pi^{3}v_{F}^{2}}\int
d^{2}p
e^{i\textbf{p}\textbf{r}}\Lambda(\textbf{p})\Gamma(\omega,\textbf{p}),\label{19}
\end{eqnarray}
with
\begin{equation}
\Gamma(\omega,\textbf{p})=\int d^{2}q \frac{8\omega^{3}-2\omega
q^{2}}{(\omega^{2}-q^{2}+i\delta)(\omega^{2}-(q-p)^{2}-i\delta)}\;,\nonumber
\end{equation}
and
$$
\Lambda(\textbf{p})=\sum_{i=1}^{N}\frac{e^{i\textbf{px}_{i}}}{p^{2}}.
$$
The correction to the local density of states comes from  the imaginary part of
equation (\ref{19}). This yields:
\begin{equation}
\delta
N(\omega,\textbf{r})=\frac{\mu}{2\pi^{3}v_{F}^{2}}2\omega\Sigma_{i=1}^{N}
\int dk\frac{J_{0}(\omega
r_{i})}{k^{2}}(4\omega^{2}-k^{2})F(\omega,k),\label{24}
\end{equation}
where the function $F(\omega,k)$ is
\begin{equation}
F(\omega,r)=\left\{\begin{array}{c}
                     \frac{-atanh(\frac{\sqrt{k^{2}-4\omega^{2}}}{k})}
                     {\sqrt{k^{2}-4\omega^{2}}},4\omega^{2}<k^{2} \\
                     \frac{atan({\frac{k}{\sqrt{4\omega^2-k^2}}})}
                     {\sqrt{4\omega^2-k^2}},4\omega^{2}>k^{2}
                   \end{array}\right\}.\nonumber
\end{equation}

\bibliography{pentagonb}
\end{document}